\documentclass[conference]{IEEEtran}
\usepackage[T1]{fontenc}
\usepackage{cite}
\ifCLASSINFOpdf
   \usepackage[pdftex]{graphicx}
\else
   \usepackage[dvips]{graphicx}
\fi
\usepackage[caption=false,font=footnotesize]{subfig}
\usepackage{amsmath}
\usepackage[cmintegrals]{newtxmath}
\usepackage{bm}
\usepackage{xcolor}

\usepackage{algorithmic}
\usepackage{algorithm}

\usepackage{url}
\begin{document}

\title{Computation Resource Allocation for Heterogeneous Time-Critical IoT Services in MEC}

\author{\IEEEauthorblockN{Jianhui Liu, Qi Zhang}
	\IEEEauthorblockA{DIGIT, Department of Engineering, Aarhus University, Denmark\\
		Email: \{jianhui.liu, qz\}@eng.au.dk}
}


%



\maketitle

\begin{abstract}
Mobile edge computing (MEC) is one of the promising solutions to process computational-intensive tasks within short latency for emerging Internet-of-Things (IoT) use cases, e.g., virtual reality (VR), augmented reality (AR), autonomous vehicle. Due to the coexistence of heterogeneous services in MEC system, the task arrival interval and required execution time can vary depending on services. It is challenging to schedule computation resource for the services with stochastic arrivals and runtime at an edge server (ES). In this paper, we propose a flexible computation offloading framework among users and ESs. Based on the framework, we propose a Lyapunov-based algorithm to dynamically allocate computation resource for heterogeneous time-critical services at the ES. The proposed algorithm minimizes the average timeout probability without any prior knowledge on task arrival process and required runtime. The numerical results show that, compared with the standard queuing models used at ES, the proposed algorithm achieves at least 35\% reduction of the timeout probability, and approximated utilization efficiency of computation resource to non-cause queuing model under various scenarios.
\end{abstract}

\IEEEpeerreviewmaketitle

\section{Introduction}
A variety of emerging Internet-of-Things (IoT) use cases, e.g., virtual reality (VR), augmented reality (AR), autonomous vehicle, factory automation, and remote surgery etc., require real-time control and steering of cyber physical systems. These use cases often involve intensive computation and have stringent requirements on latency at millisecond level \cite{Zhang2015}\cite{zhang2018towards}. However, due to limited computation capability, it is difficult for mobile devices (MDs) to complete services within the latency constraint merely by local processing. The mobile edge computing (MEC) is one of the promising solutions for time-critical services \cite{Liu2018}\cite{liu2019code}. Compared to the core cloud computing, MEC paradigm facilitates MDs to obtain powerful computation resources in the vicinity and to complete computation tasks within a low latency.

\begin{figure*}[!t]
	\begin{minipage}{0.3\textwidth}
		\centering
		\includegraphics[width=0.9\textwidth]{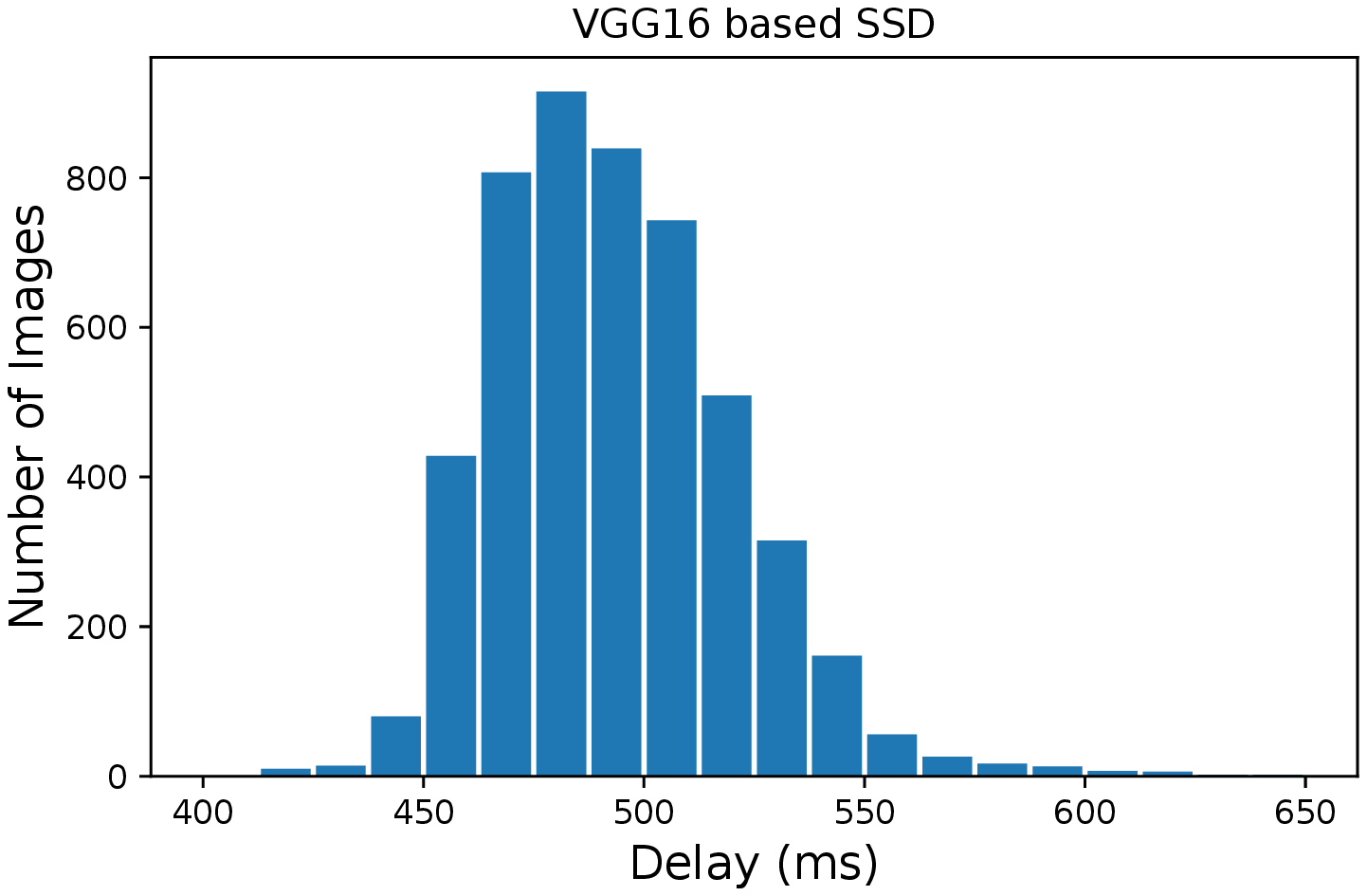}
		\caption{Histogram of object detection delay by vgg16-based SSD.}
		\label{hist_delay}
	\end{minipage}\hfill
	\begin{minipage}{0.4\textwidth}
		\centering
		\includegraphics[width=0.9\textwidth]{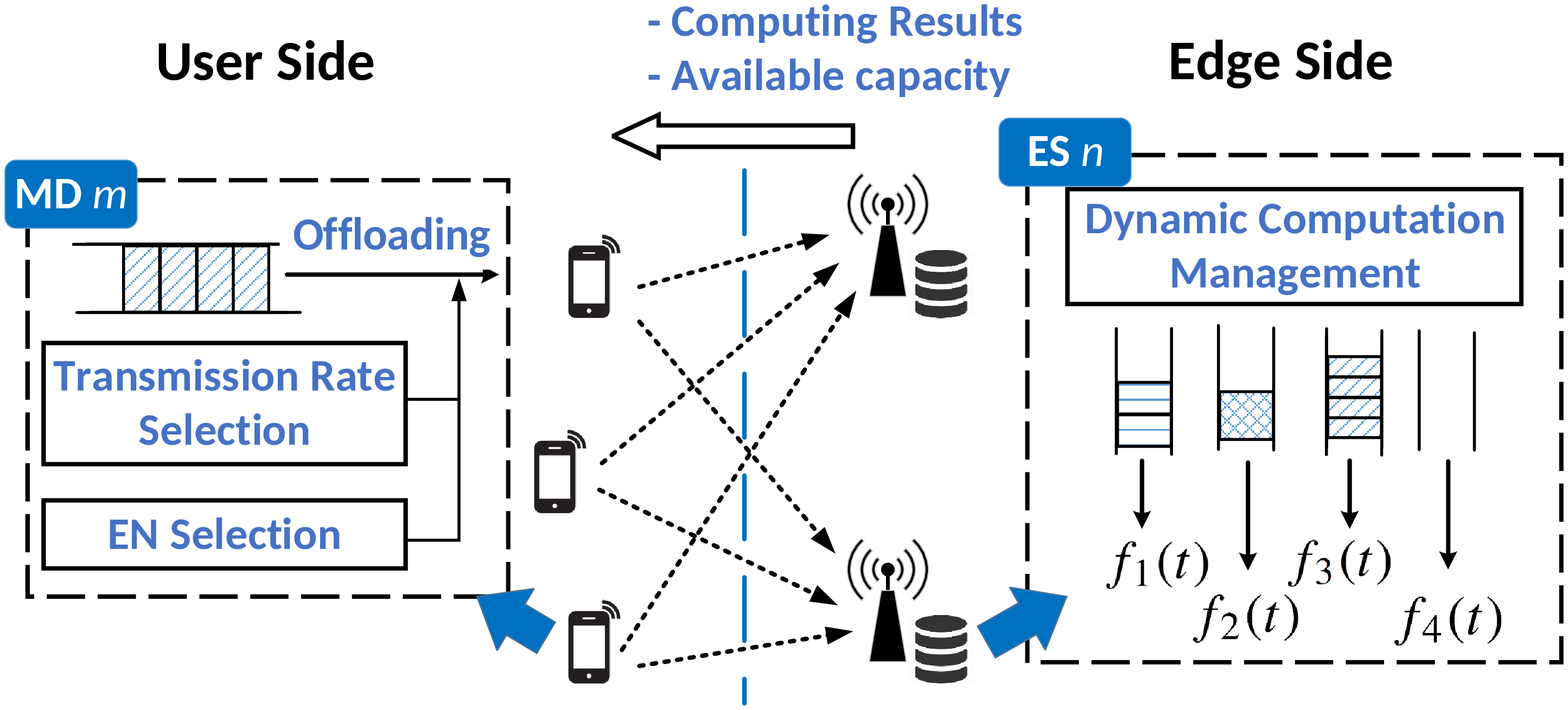}
		\caption{Proposed computation offloading framework.}
		\label{fig_off_frame}
	\end{minipage}\hfill
	\begin{minipage}{0.3\textwidth}
		\centering
		\includegraphics[width=0.9\textwidth]{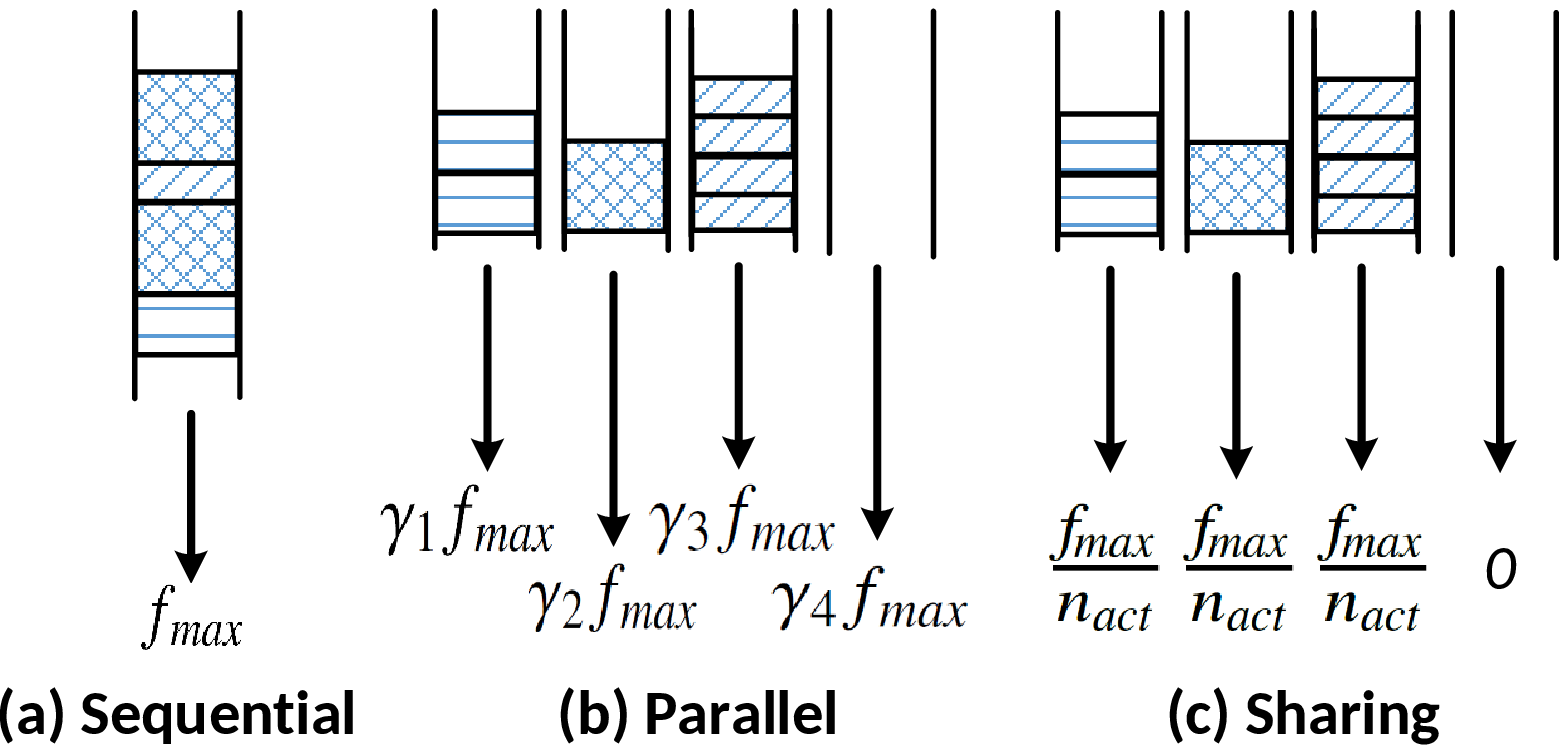}
		\caption{Standard queuing models ($\gamma_k$: ratio of average queue-k traffic load to total traffic load; $n_\textit{act}$: the number of inactive queues at the current system).}
		\label{fig_queue_model}
	\end{minipage}
\end{figure*}

Although MEC paradigm has potential to reduce the service latency, it is a challenge to fulfill the stringent latency requirement of each computation task, in particular, for the case that heterogeneous time-critical services are offloaded to one edge server (ES) from multiple MDs. First of all, the runtime of one computation task varies with different services. Even for a given service, the runtime is not deterministic. Figure \ref{hist_delay} presents the latency to detect objects on an image by Single Shot MultiBox Detector (SSD), which has no linear relationship with image size. The non-deterministic runtime makes it difficult to determine the required computing resource for each service. Moreover, each task can have different latency constraint. The reason lies in that besides heterogeneous services with different latency, the delay budget for computation depends on the time spent on transmission between MD and ES, even for a single service. Last but not least, each service has different task arrival process. For instance, the arrival of object detection for AR service is deterministic, since the image is captured at a certain rate, while other services, such as indoor labeling and annotation are in burst \cite{zhang2017networking}.

Computation offloading for MEC paradigm has attracted increasing attention in recent years. The state-of-the-art work \cite{lyu2018energy, mao2017stochastic} studied the allocation of communication and computation resource in MEC system under two impractical assumptions: (i) they assume MDs and ESs are synchronized, where the tasks of different MDs are offloaded at the same time; (ii) the computation intensity, i.e., required CPU cycles per data bits, is fixed for any task. 
Lyu \textit{et al.} \cite{lyu2018energy} jointly optimized the offloading decisions and resource allocations to guarantee task delays and energy saving. Mao \textit{et al.} \cite{mao2017stochastic} studied a joint radio and computation resource management for multi-user MEC systems, and minimized the long-term average weighted energy consumption. The algorithm failed to consider the latency constraints to complete the computation tasks.

Considering stochastic task arrival and runtime, the research work in \cite{zhao2017tasks,park2018successful,duan2018delay} allocated resource in MEC system based on queuing theory. Zhao \textit{et al.} \cite{zhao2017tasks} optimized the offloading decision and computation resource allocation to maximize the probability that tasks can meet the delay requirements. It assumed all the users have the same service, and ES processes tasks of different users following a parallel queuing model. Park \textit{et al.} \cite{park2018successful} derived the successful edge computing probability considering multi-tier network and heterogeneous tasks. The ES is assumed to process different types of tasks with a sequential queuing model. Duan \textit{et al.} \cite{duan2018delay} derived the closed-form expression of delay distribution with a sharing queuing model to minimize average latency for services with small bit sizes. They did not consider the latency constraint of each service.
The existing methods based on queuing theory results in fixed resource allocation for computation offloading, which is not efficient enough for the services with heterogeneous runtime, latency requirements and task arrival process. In addition, the resource allocation derived based on typical queuing model, e.g. M/M/n and M/D/1, is not applicable to the services with non-deterministic computation intensity.

In this paper, we study dynamic computation resource management in MEC for heterogeneous services. Considering stochastic task arrivals and computation intensity, we propose a computation offloading framework that allows offloading decisions of MDs and computation management of ESs to be optimized independently. Based on the framework, we analyze different queuing models at ES, and propose a Lyapunov optimization \cite{mao2017stochastic} based dynamic computation resource allocation algorithm for heterogeneous tasks. The proposed algorithm minimizes the average timeout probability by efficient resource management and selective task dropping, which does not require any prior knowledge on task arrival process and required runtime.  

The rest of the paper is organized as follows. Section II describes the system model and problem formulation. The proposed dynamic computation resource allocation algorithm for MEC is described in Section III. Sections IV presents the simulation results. Finally, Section V concludes the paper.

\section{System Model}
In this section, we present a computation offloading framework for MEC system and formulate an optimization problem for computation resource allocation at the ES.

\subsection{Computation Offloading Framework}

As illustrated in Fig. \ref{fig_off_frame}, we consider a network that ESs serve a set of MDs $\mathcal{M}$ ($\mathcal{M} = \{1,2,\dots,M\}$). MDs are randomly distributed and running heterogeneous services which may have different arrival processes, task sizes, required CPU cycles and latency thresholds. The average task arrival rate (in $s^{-1}$) of the MD $m$ is denoted by $\lambda_m$. $\delta_m$ denotes the latency threshold (in \textit{seconds}) to complete the task of MD $m$.
We assume there are $K$ types of tasks in the network and each MD has one type of tasks. 
The task size (in \textit{bits}) of each type is considered as a constant, such as the captured frames size in AR. However, the required CPU cycles of a task of type $k$ ($1\le k \le K$) is assumed to follow uniform distribution with the mean of $\bar{c}_k$, as it may vary with tasks.

At the user side, to shorten computation latency, MDs can offload tasks to ESs via wireless channel. As different types of tasks have different arrival process and latency constraints, it is preferable to allow MDs to make offloading decision in a distributed way according to its current channel state and requirements upon new arrival task. For example, each MD can dynamically select optimal transmission rate and the ES to which offloading the tasks, to minimize the service failure probability, which is detailed in our previous work \cite{liu2019ar}.

At the edge side, heterogeneous tasks from multiple MDs randomly arrive at one ES. The sojourn time of each task at the ES, which is the difference between the arrival and departure instant of a task, depends not only on its required CPU cycles, but also on the queuing system at the ES. As shown in Fig. \ref{fig_queue_model}, there are three standard queuing models used in the state-of-the-art work. 
The \textit{sequential queuing model} \cite{park2018successful} computes the tasks with the maximal computing capacity of the ES following the order of task arrival, which can make the most of the computation resource at the ES\cite{she2018joint}. However, if the required CPU cycles for some types of tasks follow a heavy-tailed distribution, it may lead to long waiting time for small tasks arrived after the large task.
The \textit{parallel queuing model} \cite{zhao2017tasks} maintains a queue for each type of task. The computation resource is proportionally distributed according to the ratio of average traffic load in each queue to the total traffic at the ES. Obviously, a fixed resource allocation for each queue is inefficient as a queue may become temporarily empty.
To improve efficiency, the \textit{sharing queuing model} \cite{duan2018delay} equally assigns the computing capacity of the ES to the current active queues, i.e. non-empty queues, which can uniformly distribute the computation resource to each types of tasks. However, it will be difficult to fulfill diverse latency requirements of heterogeneous tasks simultaneously.  

To meet latency constraints of heterogeneous services, we propose a \textit{dynamic queuing model}, which dynamically allocates the computation resource over time, taking into account the current queue length and tasks' waiting time. 

\subsection{Problem Formulation}
 We consider the time is slotted at an ES and indexed by $t$, where the slot length is $\tau$. For a given ES, the computation resource allocation is updated for every time slot. The ES maintains a queue buffer for each type of task in the network. The backlog of the queue $k$ ($1 \le k \le K$) at time slot $t$ is denoted by $Q_k(t)$ (in \textit{cycles}) and is evolved by 
\begin{equation}
 \label{eq_enq}
 Q_k(t+1) = \max\left\lbrace Q_k(t) + A_k(t) - \tau f_k(t),\  0\right\rbrace,
\end{equation}
where $f_k(t)$ is the computation resource that the ES assigns to queue $k$ at slot $t$. Given the maximal computation resource of ES, $f_\textit{max}$, we have $\sum_{k=1}^{K} f_k(t) \le f_\textit{max}$ for $\forall t\ge0$.
$A_k(t)$ is the sum of the required CPU cycles of the arrived tasks in queue $k$ during time slot $t$. We denote $\mathcal{M}_k$ as the set of MDs having the task type $k$, and $c_m(t)$ as the required CPU cycles of the tasks from MD $m$. $A_k(t)$ can be represented as
$A_k(t) = \sum_{m\in\mathcal{M}_k} c_m(t)$ for $\forall k \in [1,K]$ and $t \ge 0$.
Note that $A_k(t)$ cannot be accurately obtained in practice, as the required CPU cycles of each task is a random variable. Therefore, we estimate $A_k(t)$ as $\tilde{A}_k(t) = \lambda_k\bar{c}_k\tau$, where $\lambda_k$ is the arrival rate of task type $k$ at the ES.

The maximal allowed sojourn time of an arrived task from MD $m$, $\delta_m^\textit{cmp}$, is constrained by $\delta_m - T_m^\textit{tran}$, where $T_m^\textit{tran}$ is the transmission latency of a task of MD $m$. Due to stochastic channel state, $\delta_m^\textit{cmp}$ varies with $T_m^\textit{tran}$.
Therefore, denoting $\bar{T}_k$ as the average sojourn time of the tasks in queue $k$, the weighted average timeout probability to complete computation tasks at the ES is expressed as
\begin{equation}
	\bar{F}^\textit{to} = \sum_{k=1}^{K}\frac{\lambda_k}{\sum_{k=1}^{K}\lambda_k}\mathbb{P}\left\lbrace \bar{T}_k > \bar{\delta}_k^\textit{cmp} \right\rbrace,
\end{equation} 
where $\bar{\delta}_k^\textit{cmp}$ is the average threshold of sojourn time for task type $k$ and $ \bar{\delta}_k^\textit{cmp} = \mathbb{E}\left\lbrace \delta_m^\textit{cmp}, m \in \mathcal{M}_k \right\rbrace $.

Our objective is to minimize the weighted average timeout probability $\bar{F}^\textit{to}$. However, as the dynamic queuing system follows $K$ independent $G/G/1$ queue models and the service rate of each queue varies over time, it is difficult to derive a closed-form approximation of $\bar{T}_k$. Therefore, we estimate the sojourn time of the new arrived tasks in queue $k$ in time $t$, $T_k(t)$, based on the Little's Law, i.e.
\begin{equation}
\label{eq_delay_proc}
T_k(t) = \frac{Q_k(t+1)}{\overline{a}_k(t)},
\end{equation}
where $\overline{a}_k(t) = \frac{1}{t}\sum_{j=1}^{t}\sum_{m\in\mathcal{M}_k}c_m(j)$, and it denotes the average required CPU cycles of the arrived traffic (in cycles/s) in queue $k$ over $t$ time slots. In this way, the optimization problem of computation resource management can be formulated as
\begin{subequations}
	\begin{align}
	\mbox{\textbf{P1}:}\ \min\limits_{f_k(t)} &\ \ \lim\limits_{T \leftarrow +\infty}\frac{1}{T}\sum_{t=1}^{T} \sum_{k=1}^{K} \mathbb{I}\left\lbrace  T_k(t) - \bar{\delta}_k^\textit{cmp} \right\rbrace    \nonumber \\
	\textit{s.t.} &\ \ \sum_{k=1}^{K}f_k(t) \le f_\textit{max}, \ \forall t \in [1, T], \label{Prob-1-a}
	\end{align}
\end{subequations}
where $\mathbb{I}\{x\}$ is an indicator function. If $x>0$, $\mathbb{I}\{x\} = 1$; otherwise, $\mathbb{I}\{x\}=0$.
The constraint \eqref{Prob-1-a} introduces the computation resource limit of the ES at each time slot.

The objective of the problem \textbf{P1} is to minimize the weighted average timeout probability over long time. The state of the queuing system, i.e. queue length, is time-varying due to stochastic task arrivals. The computation resource at the ES has to be dynamically allocated to each queue depending on the current system state. As Lyapunov optimization is often used to optimally control a dynamic system and ensure system stability \cite{mao2017stochastic}, we propose a Lyapunov optimization based dynamic computation resource allocation algorithm to solve the problem, which is elaborated in the following section.

\section{Computation Resource Optimization in MEC}
In this section, we describe how to dynamically schedule computation resource at the ES, and solve the problem \textbf{P1} based on Lyapunov optimization.

Because the queue length is temporally correlated, the computation resource allocation is time-dependent. Since the Lyapunov optimization technique requires i.i.d. allocations in different time slots, it cannot be utilized directly here. To solve this problem, we reformulate the problem and introduce a set of non-negative parameters $\bm{q}$ (in \textit{cycles}), where $\bm{q} = \left[ q_1, q_2, \cdots, q_K\right]$. The goal is to let the length of queue $k$, $Q_k(t)$, be stabilized around $q_k$ at each slot. On the one hand, it will allow i.i.d. resource allocation over time. On the other hand, it is beneficial to reduce consecutive task timeout at the same queue due to long queue length. Denoting $\bm{Q}(t) = \left[ Q_1(t), Q_2(t), \cdots, Q_k(t)\right]$, the approximation of $\bm{Q}(t)$ to $\bm{q}$ can be measured by a non-negative Lyapunov function, i.e.
\begin{equation}
\label{eq_ly_func}
L\left[\bm{Q}(t) \right] = \frac{1}{2} \sum_{k=1}^{K} \left[Q_k(t) - q_k \right]^2.
\end{equation}

Therefore, to both stabilize the queue backlog and minimize the timeout probability at the ES, the conditional Lyapunov drift-plus-penalty is introduced
\begin{equation}
\label{eq_drif_penalty}
\Delta_V = \mathbb{E}\left\lbrace \left. L\left[\bm{Q}(t+1) \right] - L\left[\bm{Q}(t) \right] + V F^\textit{to}(t) \middle|\bm{Q}(t) \right. \right\rbrace,
\end{equation}
where $V$ is a non-negative penalty, and $F^\textit{to}(t) = \sum_{k=1}^{K} \mathbb{I}\left\lbrace  T_k(t) - \bar{\delta}_k^\textit{cmp} \right\rbrace$. In each time slot, the computation resource at the ES is allocated by minimizing $\Delta_V$, which is parameterized by the current system state and penalty $V$ subject to the constraint of the maximal computing capability at the ES.

Since $\left[ \max\left( u-v+w,0 \right) \right] ^2 \le u^2 + v^2 + w^2 -2u(v-w) $ for $\forall u,v,w>0$ \cite{Liu2017}, substituting \eqref{eq_enq} and \eqref{eq_ly_func} into \eqref{eq_drif_penalty}, we obtain
\begin{align}
\Delta_V & \le B + \mathbb{E}\left\lbrace \sum_{k=1}^{K} \left[ Q_k(t)-q_k\right]\tilde{A}_k(t) \right.+ \nonumber \\
&\ \ \left. \left. \sum_{k=1}^{K}\left[q_k - Q_k(t)\right]f_k(t)\tau + V F^\textit{to}(t) \middle|\bm{Q}(t) \right. \right\rbrace, \label{eq_inequality}
\end{align}
where $B = \frac{1}{2}\left[\left( \tau f_\textit{max}\right)^2 + \sum_{k=1}^{K}\left( A_{k,\textit{max}}\right) ^2 \right]$ and it has no effect on optimization performance. $A_{k,\textit{max}}$ denotes the maximal arriving traffic load of queue $k$ during one time slot. The inequality presents the upper bound of the drift-plus-penalty function in each time slot, which means the solution of the problem \textbf{P1} can be approximated by minimizing the right side of \eqref{eq_inequality}, i.e.
\begin{subequations}
	\begin{align}
	\mbox{\textbf{P2}:}\ \min\limits_{f_k(t)} &\ \ \sum_{k=1}^{K}\left\lbrace  \left[q_k - Q_k(t)\right]f_k(t)\tau + V \mathbb{I}\left[T_k(t) - \bar{\delta}_k^\textit{cmp} \right] \right\rbrace    \nonumber \\
	\textit{s.t.} &\ \ \eqref{Prob-1-a}. \nonumber
	\end{align}
\end{subequations}

Due to the existence of $\mathbb{I}(\cdot)$ in the objective function, the problem \textbf{P2} is non-convex.
Since $ \mathbb{I}\left[T_k(t) - \bar{\delta}_k^\textit{cmp} \right] = \mathbb{I}\left[\max\left\lbrace Q_k(t) + \tilde{A}_k(t) - \tau f_k(t),\  0\right\rbrace - \overline{a}_k(t)\bar{\delta}_k^\textit{cmp} \right]$, the queue with smaller $\left[ Q_k(t) + \tilde{A}_k(t)\right] $ will have smaller timeout probability in time slot $t$.
In this way, to simplify the problem, we will linearize the objective function as the formation of $\sum_{k=1}^{K} w_k(t) f_k(t) \tau$, where $w_k(t)$ is the weight of queue $k$ in time slot $t$ and $w_k(t) \equiv [q_k-Q_k(t)] + V\left[Q_k(t) + \tilde{A}_k(t)\right]$, and prioritize the resource allocation of the queue with smaller $w_k(t)$. Therefore, the problem \textbf{P2} is transformed to
\begin{subequations}
	\begin{align}
	\mbox{\textbf{P3}:}\ \min\limits_{f_k(t)} &\ \ \sum_{k=1}^{K} w_k(t) f_k(t) \tau    \nonumber \\
	\textit{s.t.} &\ \ \eqref{Prob-1-a}. \nonumber
	\end{align}
\end{subequations}
Note that larger $V$ leads to smaller timeout probability, but may also waste more computation capacity by executing timeout tasks. Thus, adjusting $V$ can make a tradeoff between the timeout probability and computation efficiency.

It is proven in \cite{neely2010dynamic} that, if initial queue $k$ meets $Q_k(1) \le Q_{k,\textit{max}}$ and $Q_{k,\textit{max}} = q_k + A_{k,\textit{max}}$, we have $Q_k(t) \le Q_{k,\textit{max}}$ for $\forall \ t\ge1$. We denote $\delta_{k,\textit{max}}^\textit{cmp}$ as the maximal allowed latency threshold for task type $k$. As the newly arrived tasks in queue $k$ are bound to timeout if the queue length exceeds $\delta_{k,\textit{max}}^\textit{cmp}f_\textit{max}$, $\delta_{k,\textit{max}}^\textit{cmp}f_\textit{max}$ can be regarded as the maximal allowed length of queue $k$. Therefore, we define $q_k = \delta_{k,\textit{max}}^\textit{cmp}f_\textit{max} - A_{k,\textit{max}}$.

To solve problem \textbf{P3}, the weight of each queue at the ES will be calculated before each time slot starts. The ES prioritizes the computation resource allocation to the queue with smaller weight, e.g. queue $i$. We assume that $\delta_{i,\textit{min}}^\textit{cmp}(t)$ is the smallest latency constraint among the tasks in the current queue $i$. If $\delta_{i,\textit{min}}^\textit{cmp}(t) \le 0$, the ES will allocate all its remaining computation resource to the queue $i$; otherwise, the assigned computation resource will be $\min\left\lbrace \frac{Q_i(t)}{\delta_{i,\textit{min}}^\textit{cmp}(t)}, f_\textit{remain}  \right\rbrace $, where $f_\textit{remain}$ denotes the remaining computation resource after the previous queues are allocated. Note that the empty queue is not assigned any computation resource. The algorithm will be terminated when all the queues are served or the computation resource at the ES runs out.

It is worth noticing that the number of cycles in a queue, i.e. $Q_k(t)$, cannot be accurately obtained in practice, as the required CPU cycles of each task is a random variable. However, the number of the tasks in the queue can be obtained, which is denoted by $n_k(t)$. Thus, we estimate the $Q_k(t)$ as $\tilde{Q}_k(t) = [n_k(t)+1]\bar{c}_k - c^\textit{sev}_k(t)$, where $c^\textit{sev}_k(t)$ is the executed cycles of the task being serviced from queue $k$.

Furthermore, the timeout probability can be decreased further by dropping the task that stays too long in the queue to be able to meet its latency constraint. Here, we evaluate the task in the head of each queue before each time slot starts. The task will be dropped if the unfinished CPU cycles cannot be completed within the latency constraint even using all the computation resource at the ES.

\section{Simulation Results}
In this section, the numerical results are presented to compare the performance of different queuing models at ES. The maximal computing capacity of the ES is assumed to $3\times10^{10}$ cycles/s. The simulation is based on real-time object detection application which is an important component of AR system. The captured video frames of MDs, i.e. computation tasks, are offloaded to the ES to complete the object detection. Three scenarios are analyzed based on different task sizes, computation intensities (CIs) and task arrival processes. We consider there are two types of tasks in the network in each scenario. The proposed dynamic computation resource allocation algorithm (Dynamic) is compared with the three standard queuing models (Sequential, Sharing and Parallel). We also evaluate the performance of dynamic queuing model without selective task dropping (Dynamic no drop) and non-causal dynamic queuing model (Dynamic non-causal) which assumes the required CPU cycles of each task are known beforehand. The default slot duration of computation allocation is 1 ms. Considering the tradeoff between the computation efficiency and timeout probability, the penalty $V$ is set to 2. Each simulation processes $10^5$ tasks.

\subsection{Different Task Sizes}
\begin{table*}[!t]
	\renewcommand{\arraystretch}{1.1}
	\caption{Timeout probability comparison for scenario of different task sizes.}
	\label{table_data_to}
	\centering
	\begin{tabular}{c c c c c | c c c c | c c c c}
		\hline
						   &\multicolumn{4}{c}{90\% traffic load, fixed CI} &\multicolumn{4}{c}{90\% traffic load, 10\% CI variation} &\multicolumn{4}{c}{90\% traffic load, 30\% CI variation}\\
		\hline
		$\left.\mbox{Type-I}\middle/\mbox{Type-II}\right.$ & 7/2 & 5/4    & 1/2    & 1/8    &  7/2   & 5/4    & 1/2    & 1/8    & 7/2    & 5/4    & 1/2    & 1/8 \\
		\hline
		Dynamic non-causal & 0.0044 & 0.0227 & 0.0366 & 0.0090 & 0.0050 & 0.0238 & 0.0369 & 0.0107 & 0.0077 & 0.0261 & 0.0402 & 0.0207 \\
		Dynamic 		   & 0.0045 & 0.0239 & 0.0375 & 0.0091 & 0.0052 & \underline{0.0253} & \underline{0.0387} & 0.0107 & 0.0082 & 0.0300 & 0.0465 & 0.0245 \\
		Dynamic no drop    & 0.0053 & 0.0267 & 0.0421 & 0.0094 & 0.0057 & 0.0268 & 0.0439 & 0.0111 & 0.0087 & 0.0313 & 0.0475 & 0.0247 \\
		Sequential         & 0.0100 & 0.0429 & 0.0623 & 0.0109 & 0.0101 & \underline{0.0437} & \underline{0.0642} & 0.0136 & 0.0131 & 0.0475 & 0.0693 & 0.0276 \\
		Sharing 		   & 0.0074 & 0.0412 & 0.0664 & 0.0198 & 0.0080 & 0.0410 & 0.0670 & 0.0198 & 0.0118 & 0.0461 & 0.0698 & 0.0308 \\
		Parallel 		   & 0.0666 & 0.1033 & 0.2015 & 0.2194 & 0.0653 & 0.1031 & 0.2021 & 0.2186 & 0.0619 & 0.1113 & 0.2077 & 0.2359 \\
		\hline
	\end{tabular}
\end{table*}

In this scenario, MDs have tasks with different data sizes, like video frames of different resolutions in AR services. Task type I and type II have frame sizes of 0.6 Mbits and 2.4 Mbits, respectively, which corresponds to the $320\times240$ and $640\times480$ resolution gray-scale frame with depth of 8 bits, respectively. The video frame is captured and offloaded with the rate of 25 frame per second (FPS). The average computation intensity to execute a task is 100 cycles/bit\footnote{The result is estimated by laptop of Intel i5-6300U 2.3GHz CPU, 8G RAM, where Object detection is based on Google TensorFlow Object Detection API with pre-trained model "ssd mobilenet v1 coco".}. The latency to complete a task, including both transmission and computation latency, should be shorter than the interval of frame capturing, i.e. 40 ms. Thus, the ranges of sojourn latency constrains\footnote{The upper bound of the latency constraint depends on the highest transmission rate on wireless channel. It will take about 1 ms and 5 ms to transmit 0.6Mbits and 2.4 Mbits data, respectively, using the highest modulation and coding scheme on LTE channel of 10 MHz. The lower bound is the runtime that executes 1.5$\bar{c}_k$ CUP cycles with all the computation resource at the ES. For instance, the average number of required cycles for 0.6 Mbits task is 60 Mcycles. The lower bound for 0.6 Mbit tasks is $\frac{1.5*60*10^6}{3*10^{10}}=3$ ms} at the ES are set to [3, 39] ms and [12, 35] ms, respectively, for the two types of tasks.

Table \ref{table_data_to} compares the timeout probability of queuing models to complete tasks under different traffic loads and CI variations. 30\% CI variation means the required CPU cycles of a task range from $0.7\bar{c}_k$ to $1.3\bar{c}_k$, whereas 90\% traffic load means the average required CPU cycles of the arrived tasks per second are equal to $0.9f_\textit{max}$. The second row of the table presents the ratio of the arriving traffic load between type-I and type-II tasks. For instance, "7/2" denotes the case that the number of MDs having type-I and type-II tasks is 14 and 4, respectively.
The table shows that the proposed dynamic queuing models have lower timeout probability than the other queuing models. For instance, when traffic load is 90\% and CI variation is 10\%, compared with the sequential queuing models, the timeout probability of the proposed algorithm is reduced nearly by 45\% and 40\% if the traffic load ratio is 5/4 and 1/2, receptively. The parallel queuing model has the worst performance, due to the resource waste on the inactive queues. Moreover, the dynamic queuing model shows higher gains to other queuing model, when difference of the traffic loads between the type-I and type-II is smaller, i.e., the cases "5/4" and "1/2". 
With the CI variation increasing, the timeout probability of all the queuing models rises slightly, since some of arriving tasks may require more CPU cycles. The higher CI variation also makes it more difficult to determine if tasks should be dropped, which increases the performance gaps between the dynamic queuing model with selective task dropping and the non-causal dynamic queuing model.


As part of the computation resource at the ES could be wasted to compute the timeout tasks, we evaluate the wasted computation resource of different queuing models in Fig. \ref{fig_data_wast}. Denoting $c^\textit{waste}$ as the sum of wasted CPU cycles executed by the ES during simulation, the wasted computation resource is defined as $\frac{c^\textit{waste}}{f_\textit{max}T_\textit{sim}}$, where $T_\textit{sim}$ is the simulation time. A smaller value implies the queuing system is more efficient. Fig. \ref{fig_data_wast} shows that, due to the selective task dropping, the proposed dynamic allocation algorithm achieves the best performance, which is close to that of non-causal dynamic queuing model. The wasted computation resource of the sequential queuing model is larger than the proposed algorithm, but much smaller than the parallel and sharing queuing models. The reason lies in that, compared to the other two standard models, the sequential queuing model always utilize all the computing capability to execute each arrived task. It minimizes the runtime and fails to consider the queuing delay, whereas the proposed dynamic queuing model optimizes both of them.

\subsection{Different Computation Intensities}
\begin{table}[!t]
	\renewcommand{\arraystretch}{1.1}
	\caption{Timeout probability comparison for scenario of different CIs.}
	\label{table_cycle_to}
	\centering
	\begin{tabular}{c c c c c}
		\hline
		&\multicolumn{4}{c}{90\% traffic load, 10\% CI variation} \\
		\hline
		$\left.\mbox{Type-I}\middle/\mbox{Type-II}\right.$ & 7/2  & 5/4 & 1/2 & 1/8 \\
		\hline
		Dynamic non-causal & 0.00148 & 0.01658 & 0.03124 & 0.00824 \\
		Dynamic 		   & \underline{0.00149} & \underline{0.01620} & 0.03130 & 0.00834 \\
		Dynamic no drop    & 0.00156 & 0.01745 & 0.03215 & 0.00885 \\
		Sequential         & 0.01008 & 0.04172 & 0.06144 & 0.01195 \\
		Sharing 		   & \underline{0.00604} & \underline{0.03598} & 0.05565 & 0.01599 \\
		Parallel 		   & 0.05697 & 0.08658 & 0.18350 & 0.21040 \\
		\hline
	\end{tabular}
\end{table}

In this subsection, the tasks have different computation intensities, considering MDs may require diverse detection algorithms for different types of objects and different requirements of detection precision.
We assume the average computation intensities of task type I and type II are 100 cycles/bit and 400 cycles/bit, respectively. The data size of each task is set to 0.6 Mbits. In addition, the ranges of the sojourn latency constraints are [3, 39] ms and [12, 39] ms for the two types of tasks, respectively. The frame captured rate is 25 FPS.
Table \ref{table_cycle_to} shows the timeout probability of different queuing models for the case of 90\% traffic load and 10\% CI variation. The dynamic queuing model with selective task dropping achieves similar performance as the non-causal dynamic queuing model. Compared to the other models, the gain of the dynamic queuing models increases, when traffic load is dominated by tasks of type-I. For example, compared with the sharing queuing models, the timeout probability is reduced nearly by 55\% and 75\% when traffic load ratios are 5/4 and 7/2, respectively.

Fig. \ref{fig_cycle_wast} compares the wasted computation resource among queuing models for the case of 90\% traffic load and 10\% CI variation. The proposed dynamic queuing model with selective task dropping achieves the least computation resource waste, of which performance is approximated to that of the non-causal queuing model.


\subsection{Different Task Arrival Processes}
\begin{table}[!t]
	\renewcommand{\arraystretch}{1.1}
	\caption{Timeout probability comparison for scenario of different task arrival processes.}
	\label{table_arrv_to}
	\centering
	\begin{tabular}{c c c c c}
		\hline
		&\multicolumn{4}{c}{90\% traffic load, 10\% CI variation} \\
		\hline
		$\left.\mbox{Type-I}\middle/\mbox{Type-II}\right.$ & 7/2  & 5/4 & 1/2 & 1/8 \\
		\hline
		Dynamic non-causal & 0.06623 & 0.07854 & 0.08014 & 0.03689 \\
		Dynamic 		   & 0.06666 & 0.08358 & \underline{0.08223} & 0.03827 \\
		Dynamic no drop    & 0.07439 & 0.09082 & 0.09567 & 0.04238 \\
		Sequential         & 0.07970 & 0.11450 & 0.13980 & 0.06026 \\
		Sharing 		   & 0.08988 & 0.12370 & \underline{0.12430} & 0.04915 \\
		Parallel 		   & 0.15410 & 0.21500 & 0.27380 & 0.26100 \\
		\hline
	\end{tabular}
\end{table}

\begin{figure*}[!t]
	\begin{minipage}{0.3\textwidth}
		\centering
		\includegraphics[width=1\textwidth]{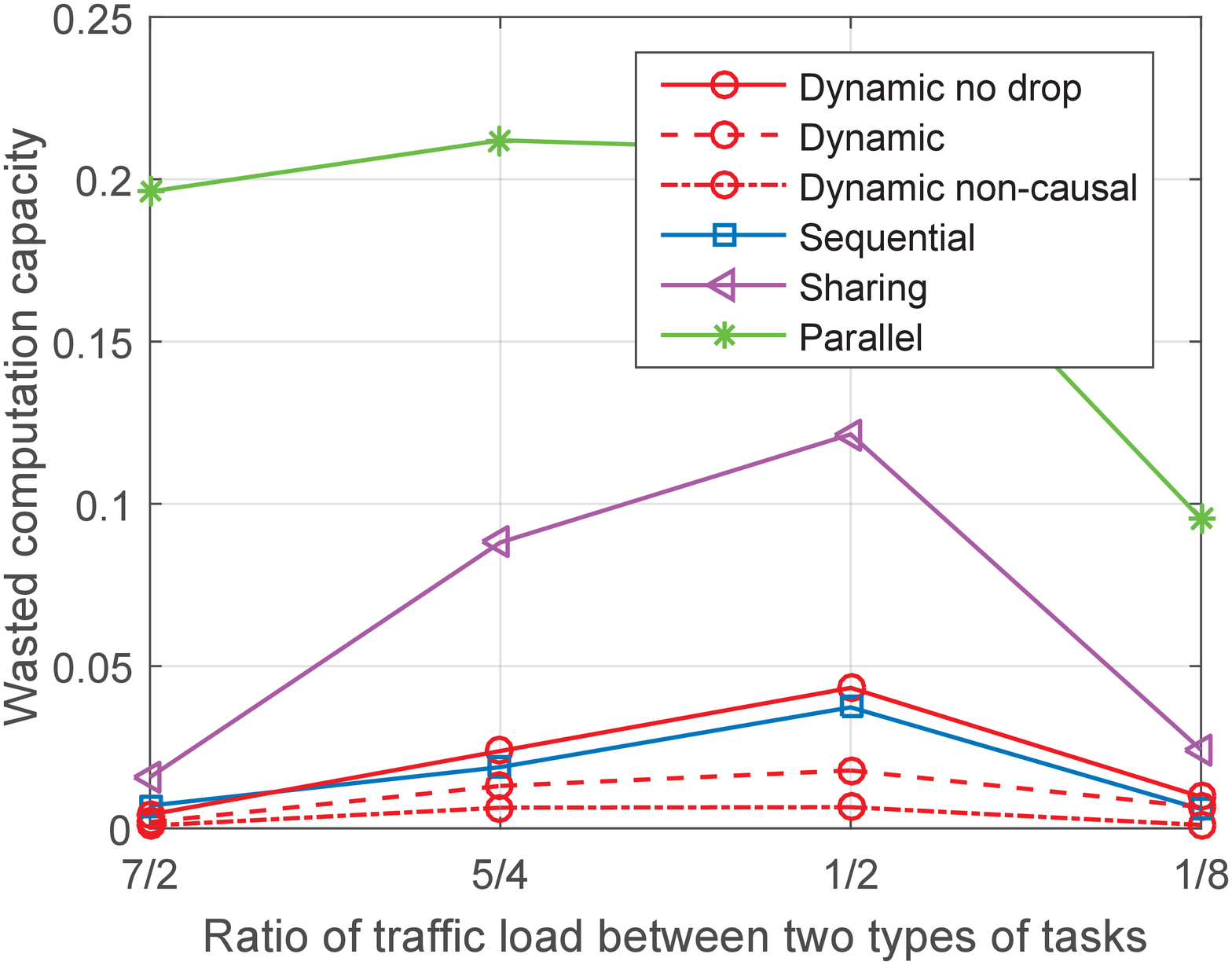}%
		\caption{Comparison of the wasted computing capacity for scenario of different task sizes: 90\% traffic load and 10\% CI variation.}
		\label{fig_data_wast}
	\end{minipage}\hfill
	\begin{minipage}{0.3\textwidth}
		\centering
		\includegraphics[width=1\textwidth]{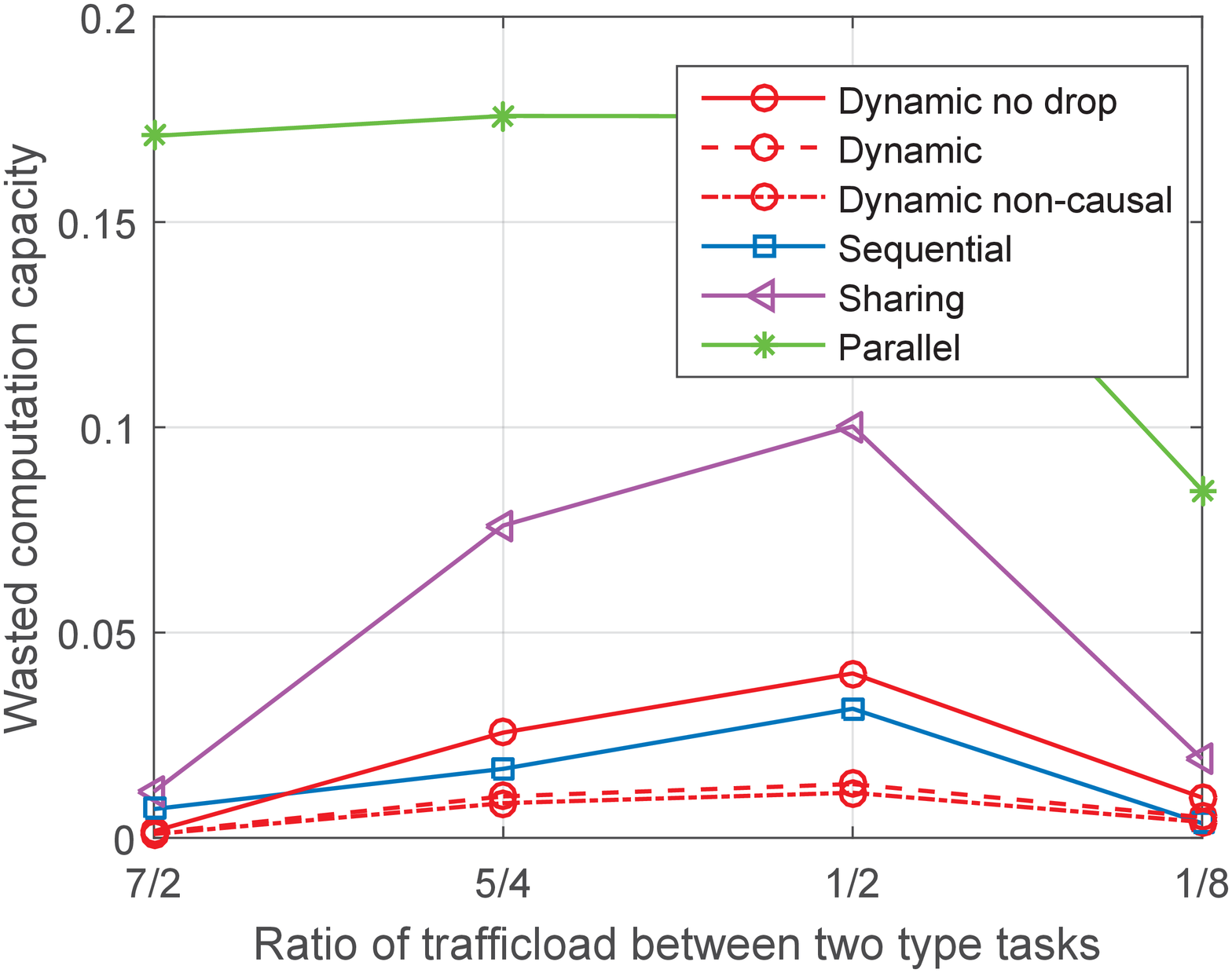}%
		\caption{Comparison of the wasted computing capacity for scenario of different CIs: 90\% traffic load and 10\% CI variation.}
		\label{fig_cycle_wast}
	\end{minipage}\hfill
	\begin{minipage}{0.3\textwidth}
		\centering
		\includegraphics[width=1\textwidth]{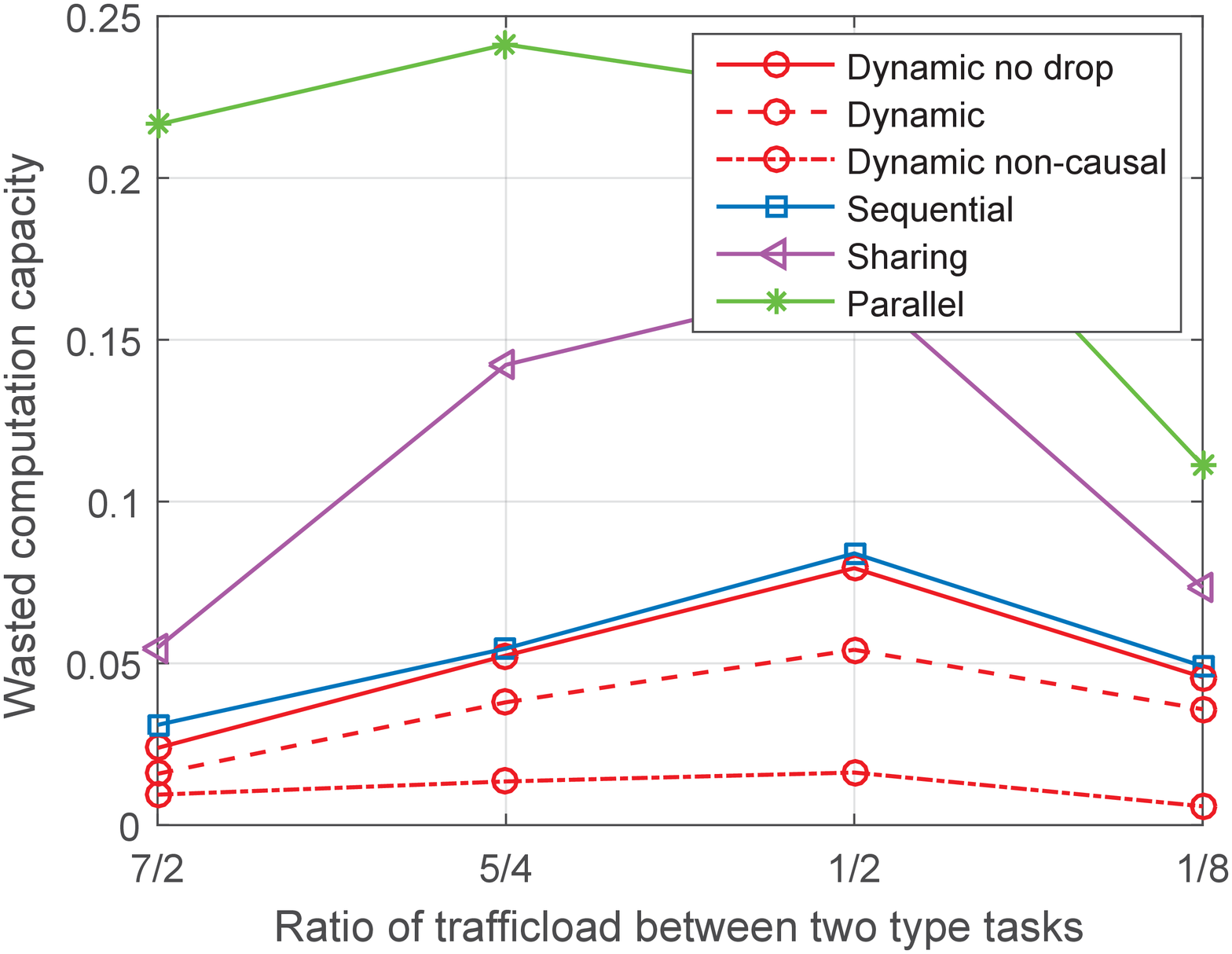}%
		\caption{Comparison of the wasted computing capacity for scenario of different arrival processes: 90\% traffic load and 10\% CI variation.}
		\label{fig_arrv_wast}
	\end{minipage}
\end{figure*}

In the previous two subsections, the arriving interval is deterministic between two consecutive tasks from the same MD. In practice, the task arrival process could be different from heterogeneous services. Therefore, in this subsection, we assume that the task arrival processes of task type-I and type-II follow the Poisson process and deterministic process, respectively. Their average task arrival rates are both 25 FPS. The other settings follow the scenario in Subsection IV-A.

Table \ref{table_arrv_to} shows the timeout probability comparison among different queuing models. The dynamic queuing model has lower timeout probability than the other queuing models. For example, compared with the sharing queuing model, the dynamic queuing model with selective task dropping reduces the timeout probability nearly by 35\%, when the traffic load ratios are 1/2. Moreover, by comparing Table \ref{table_arrv_to} and Table \ref{table_data_to}, it can be seen that different traffic arrival processes increase the timeout probability. The reason lies in that the tasks arrived in burst may lead to the traffic load temporarily over 100\%.

Fig. \ref{fig_arrv_wast} shows that both dynamic queuing models with and without selective task dropping achieve lower computation resource waste compared with the other standard queuing models.


\section{Conclusions}
In this paper, a flexible computation offloading framework is proposed for multi-user MEC system. Based on the framework, we optimize the computation resource allocation at ES considering heterogeneous time-critical services. The proposed dynamic computation allocation algorithm minimizes the weighted average timeout probability without any prior knowledge on the task arrival process and required runtime. Compared with the three standard queuing models, the proposed algorithm can achieve at least 35\% reduction of timeout probability under all the tested scenarios, and is the most efficient algorithm to utilize computation resource. The performance of the proposed algorithm is approximated to the non-causal queuing model.



\bibliographystyle{IEEEtran}
\bibliography{IEEEabrv,Ref_p4}

\begin{thebibliography}{10}
\providecommand{\url}[1]{#1}
\csname url@samestyle\endcsname
\providecommand{\newblock}{\relax}
\providecommand{\bibinfo}[2]{#2}
\providecommand{\BIBentrySTDinterwordspacing}{\spaceskip=0pt\relax}
\providecommand{\BIBentryALTinterwordstretchfactor}{4}
\providecommand{\BIBentryALTinterwordspacing}{\spaceskip=\fontdimen2\font plus
\BIBentryALTinterwordstretchfactor\fontdimen3\font minus
  \fontdimen4\font\relax}
\providecommand{\BIBforeignlanguage}[2]{{%
\expandafter\ifx\csname l@#1\endcsname\relax
\typeout{** WARNING: IEEEtran.bst: No hyphenation pattern has been}%
\typeout{** loaded for the language `#1'. Using the pattern for}%
\typeout{** the default language instead.}%
\else
\language=\csname l@#1\endcsname
\fi
#2}}
\providecommand{\BIBdecl}{\relax}
\BIBdecl

\bibitem{Zhang2015}
Q.~Zhang and F.~H. Fitzek, ``Mission critical iot communication in 5g,'' in
  \emph{Future Access Enablers of Ubiquitous and Intelligent
  Infrastructures}.\hskip 1em plus 0.5em minus 0.4em\relax Springer, 2015, pp.
  35--41.

\bibitem{zhang2018towards}
Q.~Zhang, J.~Liu, and G.~Zhao, ``Towards 5g enabled tactile robotic
  telesurgery,'' \emph{arXiv preprint arXiv:1803.03586}, 2018.

\bibitem{Liu2018}
J.~Liu and Q.~Zhang, ``Offloading schemes in mobile edge computing for
  ultra-reliable low latency communications,'' \emph{IEEE Access}, vol.~6, pp.
  2169--3536, 2018.

\bibitem{liu2019code}
------, ``Code-partitioning offloading schemes in mobile edge computing for
  augmented reality,'' \emph{IEEE Access}, vol.~7, pp. 11\,222 -- 11\,236,
  2019.

\bibitem{zhang2017networking}
W.~Zhang, B.~Han, and P.~Hui, ``On the networking challenges of mobile
  augmented reality,'' in \emph{Proceedings of the Workshop on Virtual Reality
  and Augmented Reality Network}, 2017, pp. 24--29.

\bibitem{lyu2018energy}
X.~Lyu, H.~Tian, W.~Ni, Y.~Zhang, P.~Zhang, and R.~P. Liu, ``Energy-efficient
  admission of delay-sensitive tasks for mobile edge computing,'' \emph{IEEE
  Transactions on Communications}, vol.~66, no.~6, pp. 2603--2616, 2018.

\bibitem{mao2017stochastic}
Y.~Mao, J.~Zhang, S.~Song, and K.~B. Letaief, ``Stochastic joint radio and
  computational resource management for multi-user mobile-edge computing
  systems,'' \emph{IEEE Transactions on Wireless Communications}, vol.~16,
  no.~9, pp. 5994--6009, 2017.

\bibitem{zhao2017tasks}
T.~Zhao, S.~Zhou, X.~Guo, and Z.~Niu, ``Tasks scheduling and resource
  allocation in heterogeneous cloud for delay-bounded mobile edge computing,''
  in \emph{2017 IEEE International Conference on Communications (ICC)}, 2017,
  pp. 1--7.

\bibitem{park2018successful}
C.~Park and J.~Lee, ``Successful edge computing probability analysis in
  heterogeneous networks,'' in \emph{2018 IEEE International Conference on
  Communications (ICC)}, 2018, pp. 1--6.

\bibitem{duan2018delay}
Y.~Duan, C.~She, G.~Zhao, and T.~Q. Quek, ``Delay analysis and computing
  offloading of urllc in mobile edge computing systems,'' in \emph{2018 10th
  International Conference on Wireless Communications and Signal Processing
  (WCSP)}, 2018, pp. 1--6.

\bibitem{liu2019ar}
J.~Liu and Q.~Zhang, ``Edge computing enabled mobile augmented reality with
  imperfect channel knowledge,'' in \emph{IEEE European Wireless}, 2019, pp.
  1--6.

\bibitem{she2018joint}
C.~She, C.~Yang, and T.~Q. Quek, ``Joint uplink and downlink resource
  configuration for ultra-reliable and low-latency communications,'' \emph{IEEE
  Transactions on Communications}, vol.~66, no.~5, pp. 2266--2280, 2018.

\bibitem{Liu2017}
C.-F. Liu, M.~Bennis, and H.~V. Poor, ``Latency and reliability-aware task
  offloading and resource allocation for mobile edge computing,'' in \emph{2017
  IEEE Globecom Workshops (GC Wkshps)}, 2017, pp. 1--7.

\bibitem{neely2010dynamic}
M.~J. Neely and L.~Huang, ``Dynamic product assembly and inventory control for
  maximum profit,'' in \emph{49th IEEE Conference on Decision and Control
  (CDC)}, 2010, pp. 2805--2812.

\end{thebibliography}

\end{document}